\documentclass[pre,showpacs,amsmath,amssymb,longbibliography]{revtex4-1}
\usepackage{amsmath,amssymb,latexsym,epsfig,graphics,epsf,bm}
\usepackage{graphics,xcolor}
\usepackage{float}

\newcommand{\bR}{\mathbf R}
\newcommand{\bRI}{\mathbf R^\textrm{I}}

\newcommand{\br}{\mathbf r}

\newcommand{\eq}[1]{Eq.~(\ref{#1})}
\newcommand{\Eq}[1]{Equation~(\ref{#1})}
\newcommand{\be}{\begin{equation}}
\newcommand{\ee}{\end{equation}}
\newcommand{\Ueq}{{U}_{\rm eq}}

\newcommand{\DU}{\Delta U}
\newcommand{\DT}{\Delta T}

\newcommand{\fig}[1]{Fig. ~\ref{#1}}
\newcommand{\Fig}[1]{Figure~\ref{#1}}
\newcommand{\sect}[1]{Sec.~\ref{#1}}

\begin{document}

\title{Distance-as-time in physical aging} 
\author{Ian M. Douglass}
\author{Jeppe C. Dyre}
\affiliation{{\it Glass and Time}, IMFUFA, Department of Science and Environment, Roskilde University, P. O. Box 260, DK-4000 Roskilde, Denmark} 
\date{\today}

\begin{abstract}
Although it has been known for half a century that the physical aging of glasses in experiments is described well by a linear thermal-history convolution integral over the so-called material time, the microscopic definition and interpretation of the material time remains a mystery. We propose that the material-time increase over a given time interval reflects the distance traveled by the system's particles. Different possible distance measures are discussed, starting from the standard mean-square displacement and its inherent-state version that excludes the vibrational contribution. The viewpoint adopted, which is inspired by and closely related to pioneering works of Cugliandolo and Kurchan of the 1990s, implies a ``geometric reversibility'' and a ``unique-triangle property'' characterizing the system's path in configuration space during aging. Both of these properties are inherited from equilibrium; they are confirmed by computer simulations of an aging binary Lennard-Jones system. Our simulations show that the slow particles control the material time. This motivates a ``dynamic-rigidity-percolation'' picture of physical aging. The numerical data show that the material time is dominated by the slowest particles' inherent mean-square displacement, which is conveniently quantified by the inherent harmonic mean-square displacement. This distance measure collapses data for potential-energy aging well in the sense that the normalized relaxation functions following different temperature jumps are almost the same function of the material time. Finally, the standard Tool-Narayanaswamy linear material-time convolution integral description of physical aging is derived from the assumption that when time is replaced by distance in the above sense, an aging system is described by the same expression as that of linear-response theory.
\end{abstract}

\maketitle

\section{Introduction}\label{intro}

What is the relevant measure of time during aging? This is the guiding question of the present paper. When material properties change gradually as a consequence of molecular reorganization, the term ``physical aging'' is used \cite{sim31,too31,kov63,nar71,moy76a,maz77,str78,kov79,scherer,hod95,ang00,lub04,gra06,whi06,kol12,koh13,mck17,men17,nis17,rut17,arc20,mauro,mid22}.  An example is that of a glass kept for a long time slightly below its glass-transition temperature. As pointed out by Simon long ago \cite{sim31}, such a system slowly approaches the equilibrium metastable liquid state. Physical aging is important in both production and subsequent use of three large classes of materials: covalently bonded inorganic glasses \cite{nar71,moy76a,maz77,scherer,mic16}, polymers \cite{kov63,str78,hod95,hut95,pri05,gra12,can13,mck17}, and metallic glasses \cite{che78,kho09,qia14,rut17,kuc18,lut18,son20,yiu20}. Besides these very different materials, similar aging phenomena have been reported for, e.g., spin glasses \cite{lun83,ber02}, relaxor ferroelectrics \cite{kir02}, soft glassy materials like colloids and gels \cite{fie00,fof04,pas21}, and active matter \cite{man20,jan21}.

The properties of a glass depends on the thermal history after the system fell out of equilibrium at the glass transition, and the rate of physical aging is also a function of this history \cite{har76,scherer,deb01,ber11,mck17,sca19}. From a theoretical perspective, physical aging may be regarded as an instance of response theory that considers how a given physical quantity, the ``output'', responds to an externally controlled ``input'', which in aging is usually the temperature history but also can be, e.g., the pressure history or a combination thereof. Close to the glass transition temperature, even small temperature changes result in several orders of magnitude variation of the average relaxation time. Physical aging is therefore strongly \textit{nonlinear} in the sense that property changes cannot be calculated from a linear convolution integral over the temperature variation (unless this is much smaller than one percent) \cite{scherer,mck17,arc20}. 

Examples of properties monitored in experimental studies of physical aging are: density, enthalpy, viscosity, index of diffraction, dc conductivity, frequency-dependent or  nonlinear dielectric properties, elastic moduli, and structure probed by X-rays  \cite{kov63,spi66,nar71,moy76,che78,str78,sch91,leh98,ols98,hua04,dil04,lun05,bru12,rut12,ric15}. The simplest protocol for studying physical aging is that of a temperature jump. An ideal temperature jump subjects the system to an instantaneous temperature change from a state of (metastable) thermal equilibrium, after which the system is monitored until it reaches (metastable) equilibrium at the so-called annealing temperature \cite{kov63,nar71,scherer}. The responses to both up and down jumps are generally nonexponential in time \cite{scherer}. Comparing jumps to the same temperature, a jump from above approaches equilibrium much faster and more stretched than one from below \cite{too46,kov63,scherer,mau09a,can13}. The former is referred to as self-retarded, the latter as self-accelerated \cite{can13}. This ``asymmetry of approach'' \cite{kov63,mck17} has long been understood as an effect of the so-called fictive temperature, a quantity that reflects the structure and equals the temperature for a system in thermal equilibrium \cite{too46,scherer,mck17}. The fictive temperature decreases following a down jump in temperature, which results in aging that gradually decreases. For an up jump the opposite happens \cite{kov63,scherer,hod95,mck17}. Only for very small temperature jumps does one enter into a linear regime for which up and down jumps are close to being mirror images of one another; here standard linear-response theory applies as recently demonstrated experimentally by data for jumps of down to 2 mK amplitude \cite{rie22,mid22}.  

In 1971 the Ford Motor Company engineer Narayanaswamy proposed to rationalize the asymmetry of approach by replacing the actual time $t$ by what became known as the ``material'' time $\xi$ \cite{nar71,scherer}. The beauty of Narayanaswamy's idea is that the function $\xi(t)$ embodies all nonlinear effects \cite{scherer}. If $\xi$ instead of $t$ is used as the time variable, a linear description is arrived at, i.e., almost miraculously the asymmetry between up and down jumps disappears \cite{nar71,kov79,scherer}. The formalization of Narayanaswamy's seminal discovery is nowadays referred to as the Tool-Narayanaswamy (TN) formalism \cite{too46,nar71,scherer}. It not only accounts for the above-mentioned asymmetry of approach \cite{kov63,mck17}, but also for all other generic characteristics of physical aging, e.g., the noted Ritland-Kovacs crossover effect \cite{rit56,kov63,scherer}. 

The TN formalism was devised for optimizing the cooling rate of windshields during production \cite{nar71}. A few years later, a mathematically equivalent approach was introduced for polymers \cite{kov79}, which had turned out to have aging characteristics very similar to those of oxide and other covalently bonded glasses. Because of its good description of physical aging, the TN formalism has been used in industry for decades. In academia, on the other hand, there has been only modest interest in developing and further refining this description of physical aging \cite{mck17}. That is the case even though many aspects of physical aging are still not well understood, a main challenge being to explain why the TN formalism works so well.

Aging papers often focus on identifying how the aging rate $d\xi/dt$ varies with macroscopic quantities like density, enthalpy, configurational entropy, high-frequency plateau shear modulus, etc \cite{nar71,scherer,hod97,mck17,hec19}. The present paper does not address this important problem, but focuses instead on the material-time concept itself. The TN-formalism raises a number of fundamental questions that even after 50 years remain unanswered, for instance: Why can the highly nonlinear physical-aging phenomenon be described by linear mathematics? How can the material time be related to microscopic quantities? Why is the aging of different quantities often controlled by the same material time? Why does the TN formalism work best for relatively small temperature variations? We do not provide conclusive answers to any of these questions below, but hopefully throw some light on them by connecting a number of ideas that have been around for many years.

Our main proposition is that the material time is a measure of the distance traveled of the system in configuration space. As discussed in \sect{disc}, related ``distance-as-time'' ideas have been proposed throughout the years in various contexts \cite{cug94,cha07,sch12,sch16,sch21}. To the best of our knowledge, however, a distance-as-time approach has not been related to the TN formalism. The material-time concept is known to work best for relatively small temperature variations, i.e., for systems that are only moderately perturbed from equilibrium \cite{scherer,mck17,rie22}. This is the present focus; thus we are not attempting to construct a completely general theory of physical aging. It is important to note, however, that the regime of relatively mild thermodynamic perturbations is still strongly nonlinear in the mathematical sense and, moreover, that this is the relevant regime for the modeling of many experiments, e.g., those involving a continuous cooling through the glass transition.

We find below that the standard mean-square displacement distance is not the quantity that controls the material time. The inherent distance traveled by the slowest particles, as quantified in the harmonic mean, is a significantly better candidate. Our simulations of a binary Lennard-Jones liquid demonstrate a nontrivial ``geometric reversibility'' of physical aging, which is closely related to the triangular relation between aging time-autocorrelation functions of spin-glass models that was discussed long time ago by Cugliandolo, Kurchan, and coworkers \cite{cug94,cha02,cha07}. After discussing different distance-as-time measures for the material time and comparing to simulations, we in the final part of the paper rewrite standard linear-response theory in terms of the relevant distance measure. By doing so it is demonstrated that the TN formalism follows from the \textit{Occam's razor} assumption that the very same equation applies for physical aging.

\section{Distance-as-time in thermal equilibrium}\label{eq}

This section motivates the use of displacement to define the material time of an aging system. This is done by fist discussing equilibrium results that are trivial in the sense that they are straightforward to prove rigorously. 

Consider a system of $N$ particles in volume $V$ in thermal equilibrium. The particle positions define the collective $3N$-dimensional coordinate vector $\bR\equiv(\br_1,...,\br_N)$ and the system traces out a trajectory in configuration space denoted by $\bR(t)$. For any two times $t_1$ and $t_2$ we define the distance $d_{12}$ between the configurations $\bR_1=\bR(t_1)$ and $\bR_2=\bR(t_2)$ by the mean-square displacement (MSD)

\be\label{d12_def}
d_{12}
\,=\,\frac{(\bR_2-\bR_1)^2}{N}\,.
\ee
In the thermodynamic limit ($N\to\infty$) the relative fluctuations of $(\bR_2-\bR_1)^2$ deriving from finite size effects go to zero and $d_{12}$ is a unique function of $|t_2-t_1|$; henceforth we always have this limit in mind. Although this means that no averaging is implied, we nevertheless still use the familiar term MSD. 

The above distance measure has the property that it for widely separated times is proportional to the time difference, i.e., 

\be\label{d12_limit}
d_{12}
\,\propto\, |t_2-t_1|\,\,\,\textrm{for\,\,}|t_2-t_1|\to\infty\,.
\ee
Since $d_{12}$ is a unique function of $|t_2-t_1|$, $|t_2-t_1|$ is conversely a unique function of $d_{12}$. All distance measures discussed below obey \eq{d12_limit}, which is crucial for deriving the TN equation (\sect{tn}).

For any three times $t_1<t_2<t_3$ the configurations $\bR_1$, $\bR_2$, and $\bR_3$ define a triangle in configuration space. Because the MSD increases monotonically with time, each side length determines the corresponding time difference. This implies that the triangle has the ``unique-triangle property'' \cite{dyr15} that a knowledge of any two of its side lengths determines the third. Thus if $d_{12}$ and $d_{23}$ are known, $t_2-t_1$ and $t_3-t_2$ are known, which implies a knowledge of $t_3-t_1=t_3-t_2+t_2-t_1$ and thereby of $d_{13}$. The unique-triangle property may be summarized by writing (in which $F$ is a function that may depend on the system and the thermodynamic state point)

\be\label{utp}
d_{13}
\,=\,F(d_{12},d_{23})
\,=\,F(d_{23},d_{12})\,.
\ee
The second equality sign follows from the fact that equilibrium dynamics is time reversible. 

\begin{figure}[h]
	\includegraphics[width=4cm]{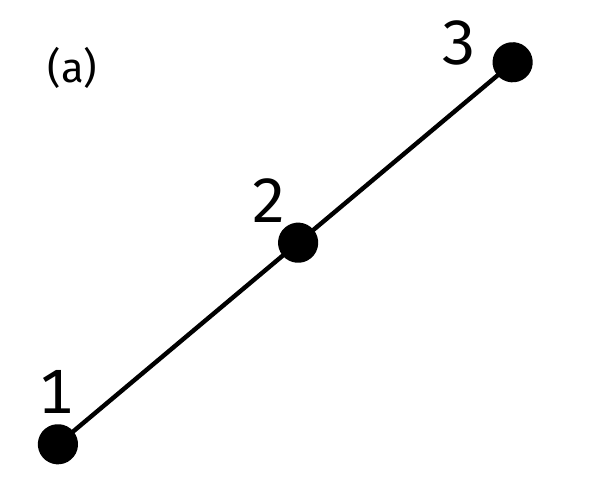}
	\includegraphics[width=4cm]{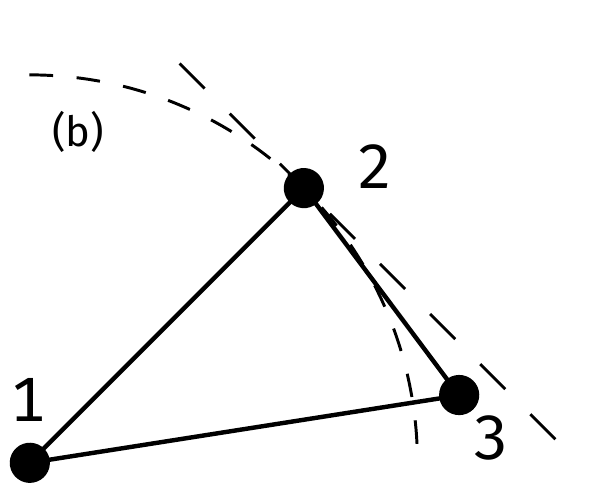}
	\includegraphics[width=4cm]{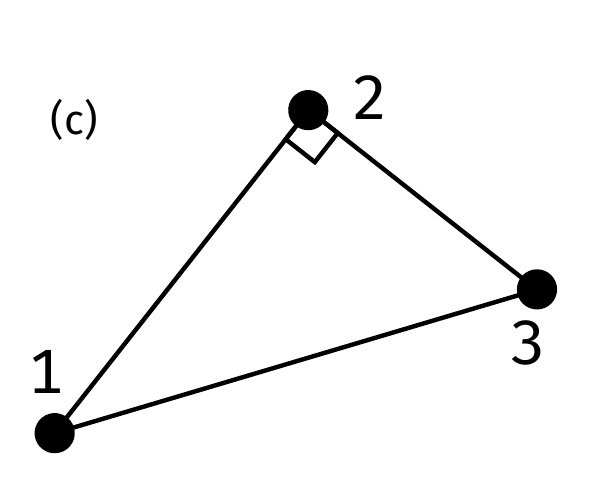}
	\includegraphics[width=5.8cm]{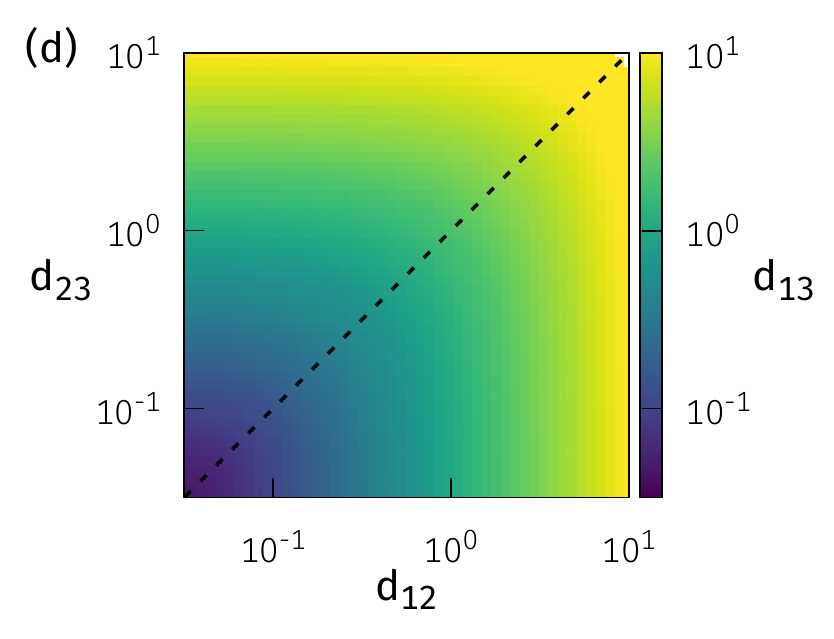}
	\includegraphics[width=5.8cm]{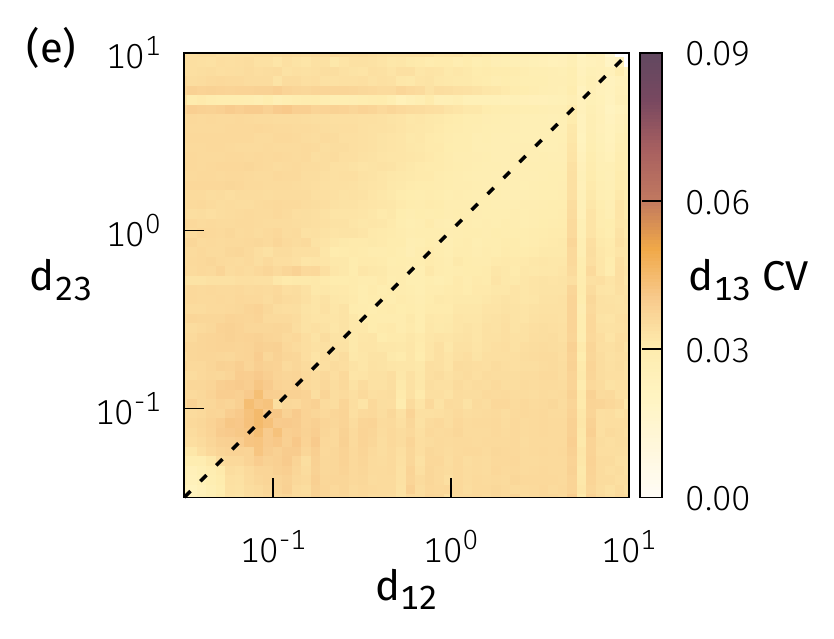}
	\includegraphics[width=5.8cm]{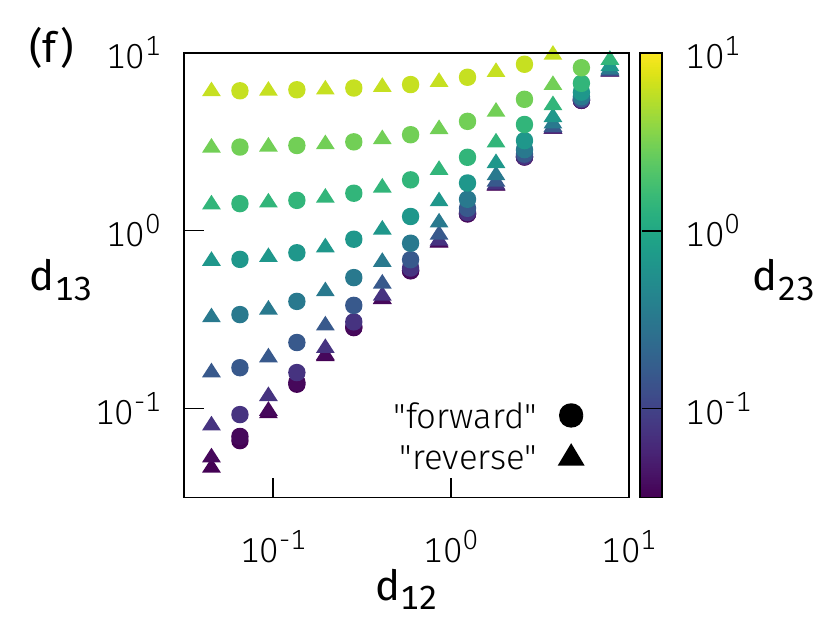}
	\caption{\label{fig1} Geometry of equilibrium dynamics.	Panels (a), (b), (c) are schematic drawings of the positions in configuration space at times $t_1<t_2<t_3$; the numerical results reported in panels (d), (e), and (f) were obtained at $T=0.60$ and density 1.2 for the modified Kob-Andersen binary Lennard-Jones system (mbLJ) \cite{sch20}. 
	(a) shows the ballistic regime, i.e., the case of very short time intervals.
	(b) illustrates the case in which the time intervals are comparable to the average relaxation time. As explained in the text, the angle between the vectors $\bR_{12}$ and $\bR_{23}$ is larger than $\pi/2$, while the length of the vector $\bR_{13}$ is larger than that of both vectors. The dashed lines mark the circle with center at $\bR_1$ and the tangent at $\bR_2$; $\bR_3$ must be between these.
	(c) shows the situation for large time intervals in which case the vectors $\bR_{12}$ and $\bR_{23}$ are uncorrelated and perpendicular. This corresponds to times at which the mean-square displacement (MSD) is linear in time.
	(d) shows the squared lengths of the vectors $\bR_{12}$, $\bR_{23}$, and $\bR_{13}$, denoted by $d_{12}$ etc, plotted in a ``heat map'' that for each value of $d_{12}$ and $d_{23}$ is colored according to the value of $d_{13}$. The diagonal symmetry is a manifestation of the time-reversal invariance of equilibrium dynamics.
	(e) A plot of the coefficient of variation (CV) (relative standard deviation) of $d_{13}$ for given values of $d_{12}$ and $d_{23}$. The fact that this quantity is small throughout shows that $d_{12}$ and $d_{23}$ determine $d_{13}$, i.e., that the unique-triangle property applies.
	(f)	shows $d_{13}$ as a function of $d_{12}$ for given values of $d_{23}$ (colors). Time reversibility implies that ``forward'' and ``reverse'' follow the same curves.}
\end{figure}

\Eq{utp} applies for all systems in thermal equilibrium and all times $t_1<t_2<t_3$. \Fig{fig1}(a) illustrates the short-time ballistic case in which $d_{13}$ is trivially determined by $d_{12}$ and $d_{23}$  because $\sqrt{d_{13}}\cong\sqrt{d_{12}}+\sqrt{d_{23}}$. A more interesting situation involves times of the same order of magnitude as the average relaxation time (b). In that case, while the length of $\bR_{13}$ is always larger than that of the two other vectors, the angle between $\bR_{12}$ and $\bR_{23}$ is larger than $\pi/2$ because the velocity autocorrelation function for any highly viscous liquid is generally negative in this region of time, leading to a negative second time derivative of the MSD. A third case (c) arises when all three time differences are much larger than the average relaxation time, which results in a right-angle triangle and $d_{13}\cong d_{12}+d_{23}$.

We proceed to present equilibrium simulation results for a Kob-Andersen type modified binary Lennard-Jones (mBLJ) liquid \cite{kob95,kob97a,kob00,sch20}. A system of 10000 particles was simulated using the GPU software RUMD \cite{RUMD}. The standard Nose-Hoover thermostat was employed with relaxation time $0.2\tau$ and time steps of $0.005\tau$ in which $\tau$ is the A particle LJ time unit. Data analysis was done using Julia by taking snapshots every 10 time steps.

\Fig{fig1}(d) shows a ``heat map'' of $d_{13}$ as a function of $d_{12}$ and $d_{23}$. As required by time reversibility, the figure is symmetrical about the $d_{12}=d_{23}$ axis (compare \eq{utp}). \Fig{fig1}(e) shows the relative standard deviation of $d_{13}$ for given values of $d_{12}$ and $d_{23}$, denoted by CV for coefficient of variation. The fluctuations are minute. This confirms the unique-triangle property which, as mentioned, always applies in equilibrium. Based on the same data, (f) shows how $d_{13}$ depends on $d_{12}$ (circles) for given values of $d_{23}$ (different colors) and how $d_{13}$ depends on $d_{23}$ for given values of $d_{12}$ (squares). The collapse is a consequence of time reversibility. These results apply rigorously so the simulations merely illustrate well-known facts.

Other distance measures than the instantaneous MSD may be used. In particular, we shall focus on measures based on the inherent dynamics \cite{sch00a}. Recall that any configuration $\bR(t)$ may be quenched to the nearest potential-energy minimum by following the steepest descent downhill. This results in the configuration's so-called inherent state  \cite{sti83}. In this way the ``inherent dynamics'' $\bRI(t)$ is defined \cite{sch00a}, in terms of which one may define the alternative distance $d_{12}^I$ between configurations at times $t_1$ and $t_2$,

\be\label{d12I_def}
d_{12}^I
\,=\,\frac{(\bRI_2-\bRI_1)^2}{N}\,.
\ee
Our simulation results for the equilibrium inherent MSD are entirely similar to those of \fig{fig1} for the equilibrium MSD by obeying geometric reversibility and the unique-triangle property (not shown). For simplicity of notation, we henceforth drop the superscript ``I'' and denote any distance measure by $d_{12}$ or just $d$.

\section{Physical aging}\label{pa}

According to the TN formalism, in a specific sense an aging material responds linearly to external stimuli like a temperature variation. Suppose the externally controlled input ``effort'' (field) is denoted by $e_b(t)$ (e.g., temperature, electric field, shear stress, ...) while the output ``charge'' is denoted by $q_a(t)$ (e.g., heat/entropy, dipole moment, shear displacement, ...). We allow for $a\neq b$, corresponding to different so-called energy bonds \cite{paynter,ost71,systemdyn,III}. The TN equation is the following Stieltjes-type convolution integral in which $\xi=\xi(t)$ is the material time, $\psi_{ab}(\xi-\xi')$ is the response function, and it is assumed that $\langle q_a\rangle=0$,

\be\label{TN_resp}
q_a(t)
\,=\,\int_{-\infty}^{\xi(t)}\psi_{ab}(\xi(t)-\xi')\,\delta e_b(\xi')\,.
\ee
One defines this kind of integral by chopping up the path in configuration space into infinitesimal pieces, each of which has the change of field $\delta e_b(\xi')$. The Stieltjes integral can be rewritten in the following more familiar form \cite{nar71,scherer} 

\be\label{TNeq}
q_a(t)
\,=\,\int_{-\infty}^{\xi(t)} \psi_{ab}(\xi(t)-\xi')\,\frac{d e_b(\xi')}{d\xi'}\,d\xi'\,.
\ee 
As mentioned, the TN description accounts for characteristics of aging like the asymmetry of approach and the Ritland-Kovacs crossover effect \cite{rit56,kov63,scherer}. The most commonly used versions of TN assume that the aging rate is a function of the property monitored \cite{nar71,scherer,moy76a,mck17,sch17,roe19}. In the simplest ``single-parameter-aging'' case, this function is an exponential \cite{hec15,hec17,roe19,rie22}. 

Previous publications focused more on applying the TN formalism than on understanding the origin of the material time. The present paper investigates the possibility of defining $\xi$ from a distance measure in configuration space. In the 1990s Kurchan and Cugliandolo proposed closely related ideas in the context of spin glass models studied by means of sophisticated theoretical techniques \cite{cug94,cha02,cha07}; these authors, however, did not discuss the connection to the TN material-time description. More recently, Schober interpreted and discussed aging simulations in terms of the mean-square displacement (MSD) \cite{sch12,sch16,sch21}. In \sect{disc} we return to the relation between these and other works and the present considerations.

Suppose the material time is controlled by a distance measure $d_{12}$ such that the difference in material time corresponding to the actual times $t_1<t_2$ is a function of $d_{12}$,

\be\label{xid12}
\xi(t_2)-\xi(t_1)
\,=\, f(d_{12})\,.
\ee
Here $d_{12}$ may be defined in terms of the MSD of \eq{d12_def}, the inherent MSD of \eq{d12I_def}, or some other distance measure. How can one in a simulation test whether \eq{xid12} is a consistent assumption? We proceed to show that \eq{xid12} implies the unique-triangle property \cite{dyr15} (the first equality sign of \eq{utp}), as well as the symmetry expressed by the second equality sign that will be referred to as ``geometric reversibility''. Consider an aging system at three times $t_1<t_2<t_3$ with corresponding material times $\xi(t_1)<\xi(t_2)<\xi(t_3)$. Then \eq{xid12} implies

\begin{eqnarray}\label{xi_eqs}
\xi(t_2)-\xi(t_1)\,&=&\,f(d_{12})\,\nonumber\\
\xi(t_3)-\xi(t_2)\,&=&\,f(d_{23})\,\\
\xi(t_3)-\xi(t_1)\,&=&\,f(d_{13})\,.\nonumber
\end{eqnarray}
Since $\xi(t_3)-\xi(t_1) =\xi(t_3)-\xi(t_2) +\xi(t_2)-\xi(t_1)$, the two distances $d_{12}$ and $d_{23}$ determine the third, $d_{13}$. This establishes the first equality sign of \eq{utp}, the unique-triangle property. A special case is that of thermal equilibrium for which the material time is proportional to the actual time. This implies that the function $F$ in \eq{utp} can be determined from equilibrium simulations. Before proceeding, we note that \eq{xi_eqs} is equivalent to stating that in terms of a ``scaling function'' $h(t)\equiv\exp(\xi(t))$, the distance can be written $d_{12}=F(h(t_1)/h(t_2))$; this formulation has been shown to apply for the random field Ising model \cite{cug94}. 

To prove the second equality sign of \eq{utp}, the symmetry of the function $F(x,y)$, we note that a time $t_4$ exists such that $\xi(t_4)-\xi(t_1)=\xi(t_3)-\xi(t_2)$. This implies $d_{14}=d_{23}$, and because $\xi(t_3)-\xi(t_4) =\xi(t_2)-\xi(t_1)$ one likewise has $d_{43}=d_{12}$. Since $\xi(t_3)-\xi(t_1) =\xi(t_3)-\xi(t_4) +\xi(t_4)-\xi(t_1)$, the distances $d_{14}$ and $d_{43}$ determine $d_{13}$, i.e., $d_{13}=F(d_{14},d_{43}) =F(d_{23},d_{12})$. Alternatively, one can derive the symmetry of $F(x,y$ simply by referring to the time reversibility of equilibrium dynamics.

Which measure of the material time should be used, the standard MSD, a measure based on the inherent particle displacements, or something else? If the temperature is changed discontinuously as in a temperature jump, the MSD changes within picoseconds because the vibrational degrees of freedom thermalize quickly. The physical idea behind the material time, however, allows only for slow and gradual changes because $\xi$ variations are assumed to reflect \textit{structural} changes \cite{nar71,scherer}. The issue associated with using the standard MSD is avoided by using an \textit{inherent} displacement measure because the inherent state does not change on the vibrational time scale. Consequently, we henceforth focus on inherent displacement measures.

\begin{figure}[h]
	\includegraphics[width=6cm]{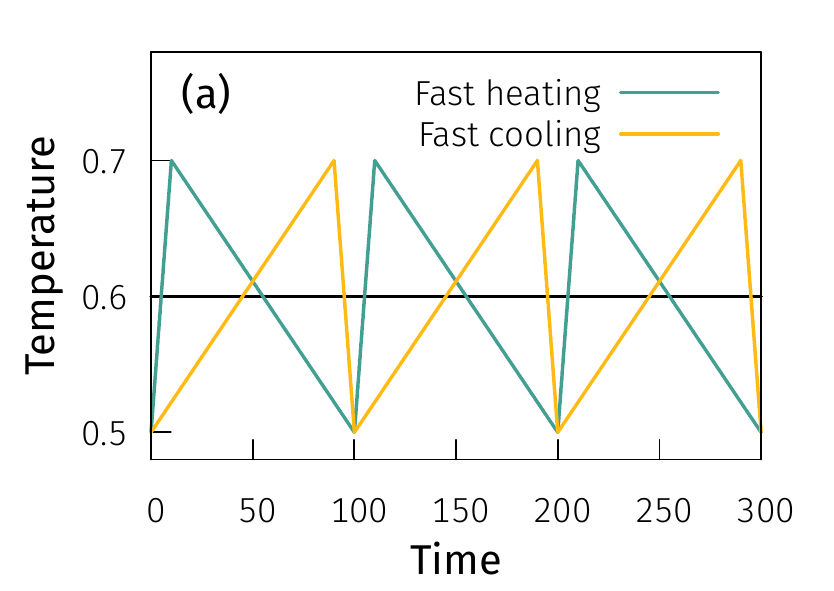}
	\includegraphics[width=6cm]{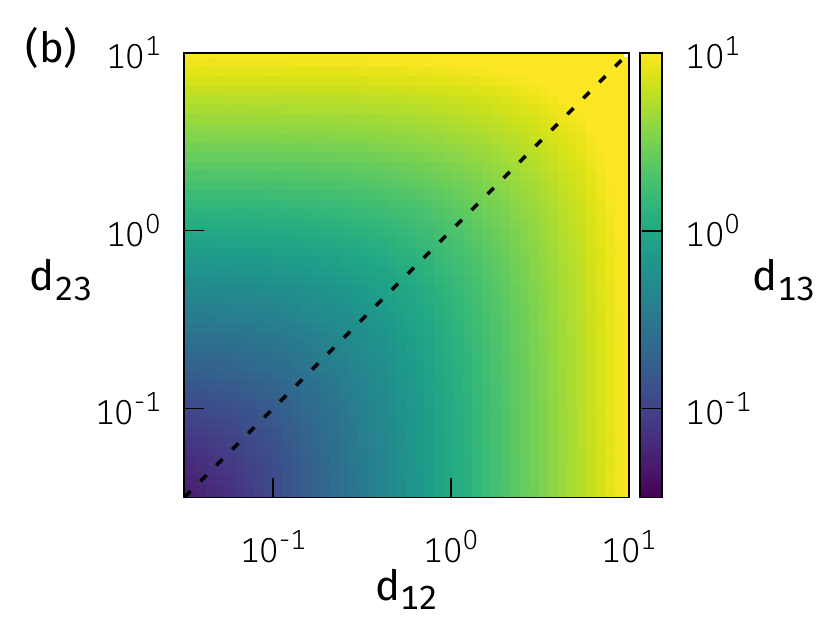}
	\includegraphics[width=6cm]{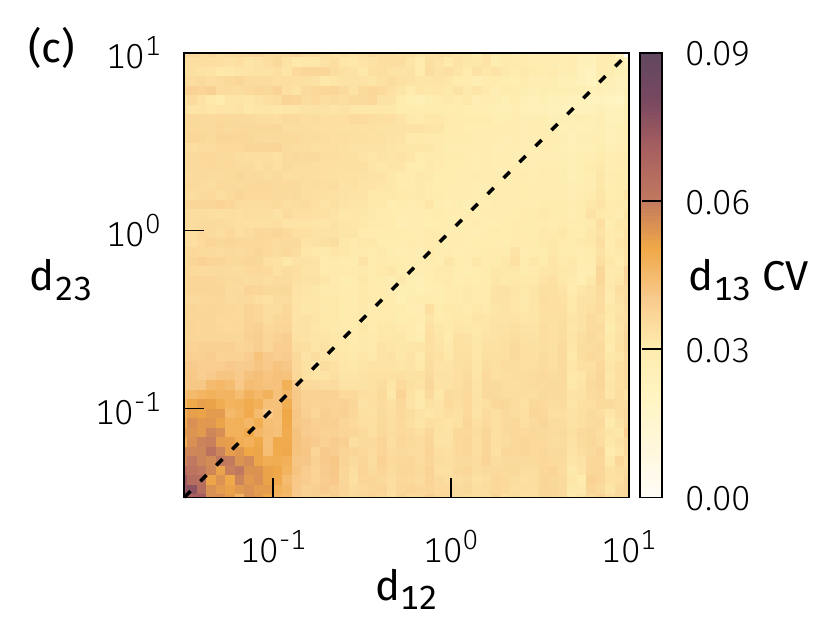}
	\includegraphics[width=6cm]{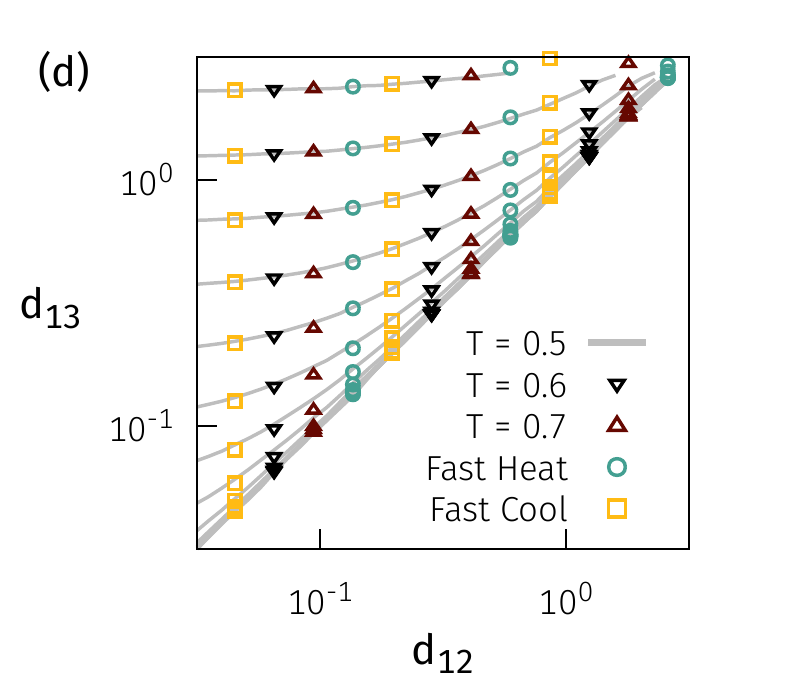}
	\caption{\label{fig2} Geometry of aging inherent dynamics. 
	(a) shows two time-asymmetric temperature protocols with, respectively, fast heating and fast cooling, working between the temperatures $0.5$ and $0.7$. The period was chosen to be comparable to the average relaxation time at $T=0.60$, implying that the system is virtually in equilibrium at the higher temperatures but essentially frozen at the lower temperatures during the cycles. 
	(b) shows a heat map like that of \fig{fig1} for the fast-heating protocol. This is not time-reversible so the observed geometric reversibility is not a given.
	(c) reports the CV for both temperature protocols. The fact that the CV is small demonstrates that that the unique-triangle property applies to a good approximation.
	(d) shows a plot like that of \fig{fig1}(f) of $d_{13}$ as a function of $d_{12}$ for given values of $d_{23}$. Data for both aging protocols of (a) are shown here together with equilibrium data at $T=0.5$, $T=0.6$, and $T=0.7$. The data collapse demonstrates the unique-triangle property that the relation between the lengths of the 1-2-3 triangle are the same during aging as in equilibrium.}
\end{figure}

Turning to physical-aging numerical data for the inherent dynamics, \fig{fig2}(a) shows two time-asymmetric cyclic temperature protocols between $T=0.50$ and $T=0.70$ with a period that is of the same order of magnitude as the average relaxation time at $T=0.60$. To get good statistics, we have averaged over many periods. \Fig{fig2}(b) shows the heat map of the inherent MSD for the fast-heating protocol. The picture is very similar to that of thermal equilibrium. In particular, the data demonstrate geometric reversibility, which is not trivial due to the irreversible nature of physical aging. Similar data are obtained for the yellow slow-heating protocol (not shown). \Fig{fig2}(c) shows the analog of \fig{fig1}(e) for both temperature protocols of \fig{fig2}(a). We see a good data collapse as indicated by a small CV. Finally, \fig{fig2}(d) shows $d_{13}$ as a function of $d_{12}$ for different fixed values of $d_{23}$. The gray lines are the equilibrium curves at $T=0.50$. Other equilibrium and aging data follow these lines. This confirms the unique-triangle property \eq{utp} for aging and shows, in particular, that the same relation between the three distances applies in equilibrium and during aging.

Having established the unique-triangle property and geometric reversibility for aging using the distance measure derived from the inherent MSD, we proceed to interpret aging data using the distance-as-time approach. The property monitored is the potential energy $U$, which plays the role of the ``charge'' on the left-hand side of \eq{TN_resp} while temperature is the externally controlled field (the ``effort''). Temperature jumps imposed at $t=0$ were studied for the mBLJ system. After a jump one monitors how $U(t)$ approaches the equilibrium value. For a jump at $t=0$ of magnitude $\DT$ from the initial temperature $T_i$ to the ``annealing'' temperature $T$, we define the normalized relaxation function $R(t)$ by

\be\label{R}
R(t)
\,\equiv\,\frac{U(t)-\Ueq(T)}{\Ueq(T_i)-\Ueq(T)}\,.
\ee
By construction, $R(t)$ goes from unity at $t=0$ to zero as $t\to\infty$. The prediction of the TN formalism \eq{TN_resp} is that $R(t)$ is the same function of the material time for all temperature jumps, i.e., that a unique function $J(\xi)$ exists such that for all jumps

\be\label{TN_jumps}
R(t)
\,=\,R_{TN}(\xi)\,.
\ee
Here $\xi$ is the increase in material time after the jump was initiated, i.e., $\xi\equiv\xi(t)-\xi(0)$, and by reference to \eq{TN_resp} $R_{TN}(\xi)\equiv\psi_{ab}(\xi)/C_V$ (in the present case of just a single energy bond, $a=b$ and the specific heat $C_V=\DU/\DT$ is assumed to be temperature independent). 

\begin{figure}[h]
	\includegraphics[width=7cm]{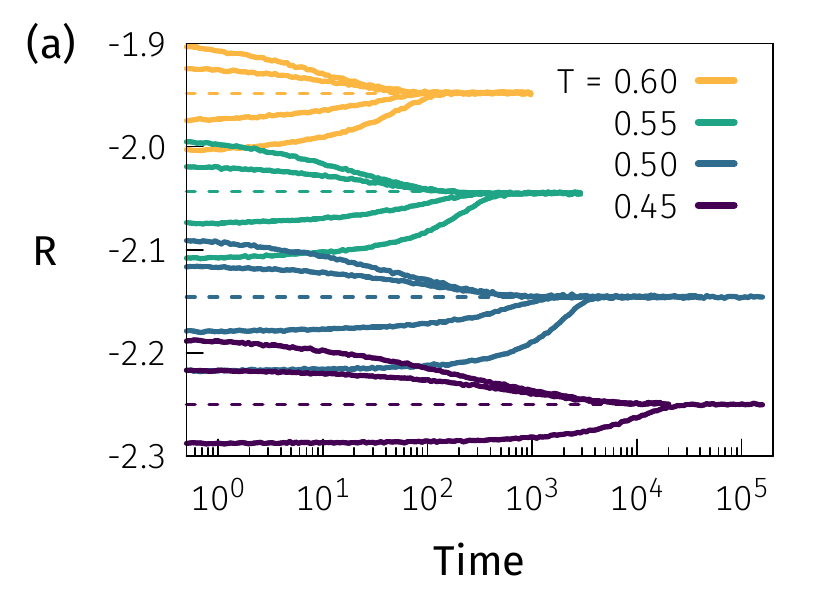}
	\includegraphics[width=7cm]{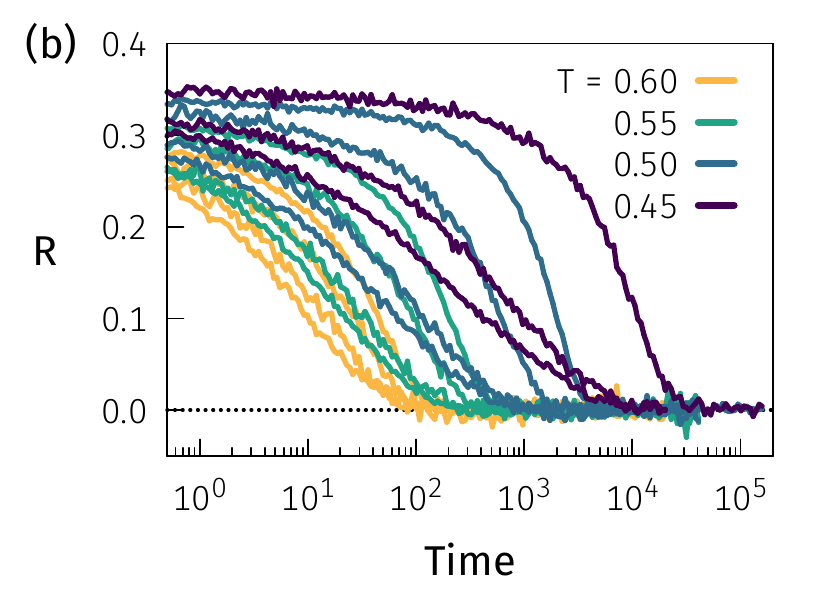}
	\includegraphics[width=7cm]{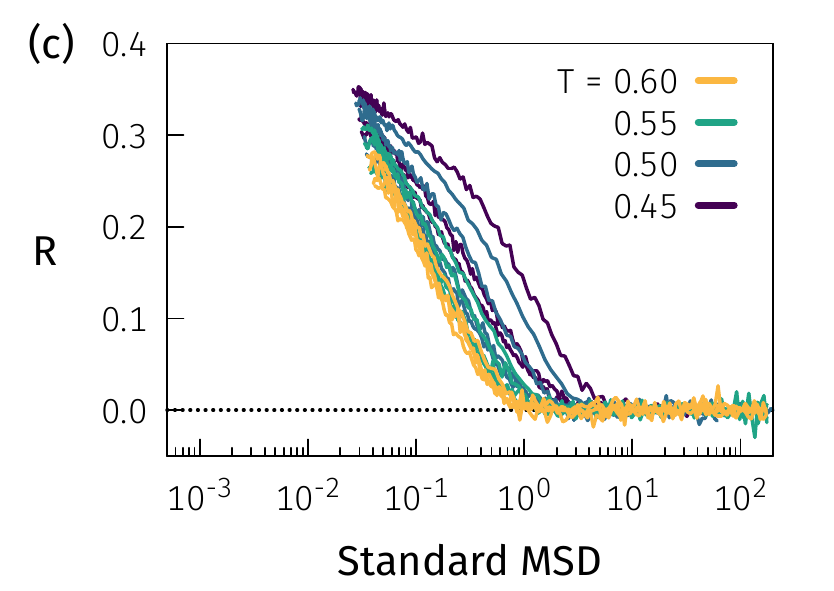}
	\includegraphics[width=7cm]{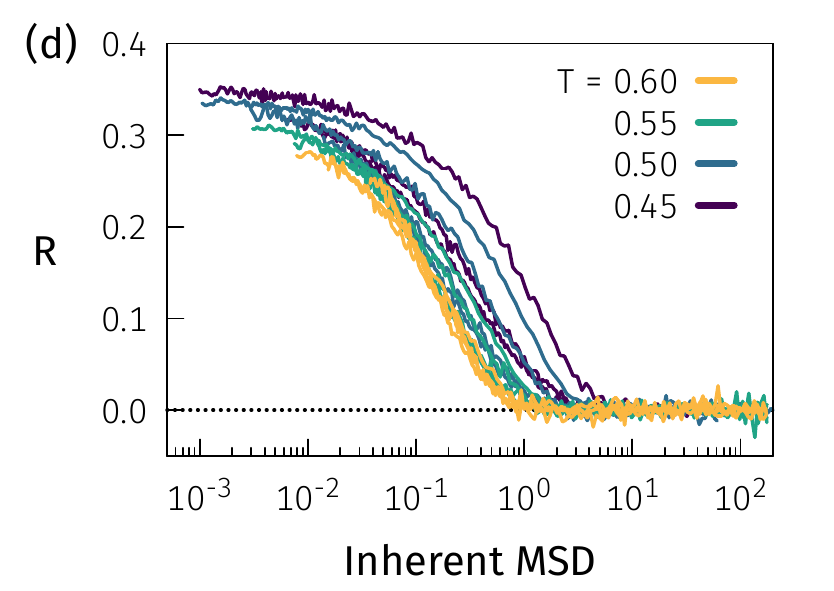}
	\caption{\label{fig3} Results from temperature-jump simulations monitoring the potential energy $U$. The color signals the annealing (final) temperature of the jump, $T$. Jumps of magnitude $\pm 0.05$ and $\pm 0.10$ were studied (no jump starts at $T_i=0.35$ due to a lack of equilibrium data). 
	(a) The potential energy as a function of the time after each jump was initiated where the dashed line represents the equilibrium value approached as $t\to\infty$.
	(b) The corresponding normalized relaxation functions $R(t)$ (\eq{R}). The data cover roughly 3.5 decades of relaxation times.
	(c) $R$ plotted as a function of the standard MSD.
	(d) $R$ plotted as a function of the inherent mean-square displacement.}
\end{figure}

\Fig{fig3}(a) presents data for $U(t)$ following jumps of magnitudes $\pm$0.05 and $\pm$0.10 to $T=$ 0.45, 0.50, 0.55, and 0.60, plotted as functions of the time after each jump was initiated. Jumps from higher temperatures thermalize more quickly than those coming from below; this is the above-mentioned ``asymmetry of approach'' reported in numerous experiments \cite{scherer,mck17}. Also as in experiments, the system takes much longer time to equilibrate if the annealing temperature $T$ is lowered; this reflects the well-known slowing down of the dynamics of glass-forming liquids, \textit{in casu} the supercooled Kob-Andersen model \cite{kob95}. \Fig{fig3}(b) shows the corresponding normalized relaxation functions $R(t)$. These do not start in unity because there is a large, almost instantaneous change of the potential energy coming from the vibrational degrees of freedom that thermalize very quickly. 

\Fig{fig3}(c) uses the standard MSD for defining the material time via \eq{d12_def} by plotting the normalized relaxation function as a function of the MSD to investigate whether the data collapse (\eq{TN_jumps}). Although the data now vary just one decade compared to the 3.5 decades in time (\fig{fig3}(b)), one cannot say that there is a good collapse. Besides, as already mentioned, a material time defined from the standard MSD jumps virtually discontinuously in a temperature jump, and for this reason alone, the standard MSD is not a good candidate for a distance-as-time measure of the material time. 

\Fig{fig3}(d) shows the same data plotted versus the inherent mean-square displacement that eliminates the unphysical vibrational jump at $t=0$. This does not provide a much better data collapse, however, although one notes a different shape of the curves at short times compared to those of \fig{fig3}(c): an approach to short-time plateaus is now visible. Nevertheless, we conclude that neither the MSD nor the IMSD provide a satisfactory measure of the material time.

\section{Role of dynamic heterogeneities}\label{dh}

This section discusses a possible explanation for why neither of the above two ``distance-as-time'' definitions of $\xi$ work well for collapsing the normalized relaxation functions. It is known that relaxation in supercooled liquids and glasses does not take place in a spatially homogeneous fashion  \cite{edi00,dynhet}. For the equilibrium viscous liquid, over time intervals of order the average relaxation time there are regions with many particle rearrangements and regions with few. After some time fast regions become slow and \textit{vice versa}. This ``dynamic heterogeneity'' has been documented in several experiments and computer simulations over the last 25 years \cite{edi00,dynhet,kar14}. Dynamic heterogeneity has also been discussed in connection with theories for physical aging \cite{pri05,cha07,mic16,sch17,lul20,pas21}. A simple way to understand how spatial heterogeneity may influence the dynamics is via elastic models of glass-forming liquids \cite{dyr06,wid06,lar08}: Because a liquid is disordered, there must be regions of varying softness \cite{sch16a,nan21}; this leads to different barriers for molecular rearrangements \cite{sch17,kap21} and, consequently, spatially varying levels of activity in the form of molecular rearrangements.

The material-time concept was devised long before dynamic heterogeneity came into focus. Attempts have been made to model physical aging in terms of a local material time with an aging rate that varies in space (and time) \cite{cha07,cas07}. We here adopt a simpler and more ``global'' path of reasoning, however. As long as a percolating structure exists of particles that have not -- or almost have not -- moved in a given time interval, the structural memory is largely maintained. This suggests that the slowest moving particles control the material time, while clusters or strings \cite{don98,glo00,sha13,hig18} of fast-moving particles are less important for structural relaxation. For instance, the movement of a string of particles in a fixed structure will not change the system's potential energy very much or relax significantly an overall shear stress. Quoting Higler \textit{et al.} \cite{hig18a}, the assumption is that ``the presence of long-lived bonded structures within the liquid may provide the long-sought connection between local structure and global dynamics''. We identify these long-lived structures as the particles that over a given time interval have moved the shortest distance in their \textit{inherent} motion. A closely related idea based on a machine-learned softness parameter was recently discussed by Schoenholz \textit{et al.} \cite{sch17}, and the importance of the slowest particles was also emphasized earlier by Szamel and coworkers \cite{kum06,sza06}. In fact, the idea that the slowest particles dominate structural relaxation is more than 30 years old, see e.g., Ref. \onlinecite{glo00} and references therein. Thus Stillinger in 1988 proposed a picture according to which a glass-forming liquid ``is viewed as a dynamic patchwork of relatively strongly bonded (but amorphous) molecular domains that are separated by irregular walls of weakened bonds'' \cite{sti88}. He argued that, as a consequence, the Stokes-Einstein relation between viscosity and diffusion coefficient (see, e.g., Ref. \onlinecite{cos19} and its references) is violated by predicting too large a diffusion coefficient because fast particles contribute a lot to the MSD, but not much to overall stress relaxation. This was subsequently confirmed \cite{fuj92,cic96,and97,yam98,edi00}.

\begin{figure}[h]
	\includegraphics[width=4.5cm]{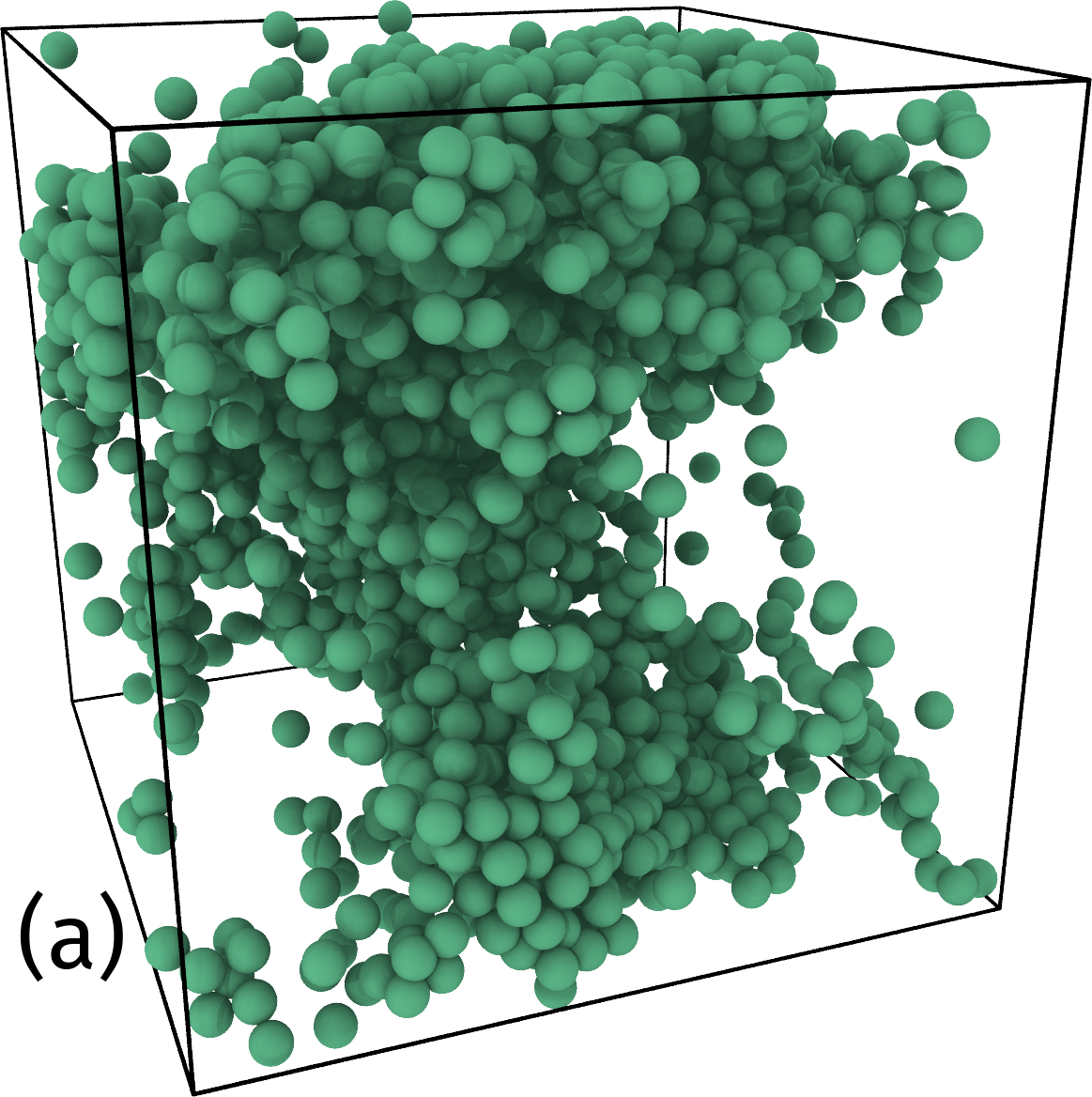}
	\includegraphics[width=6.5cm]{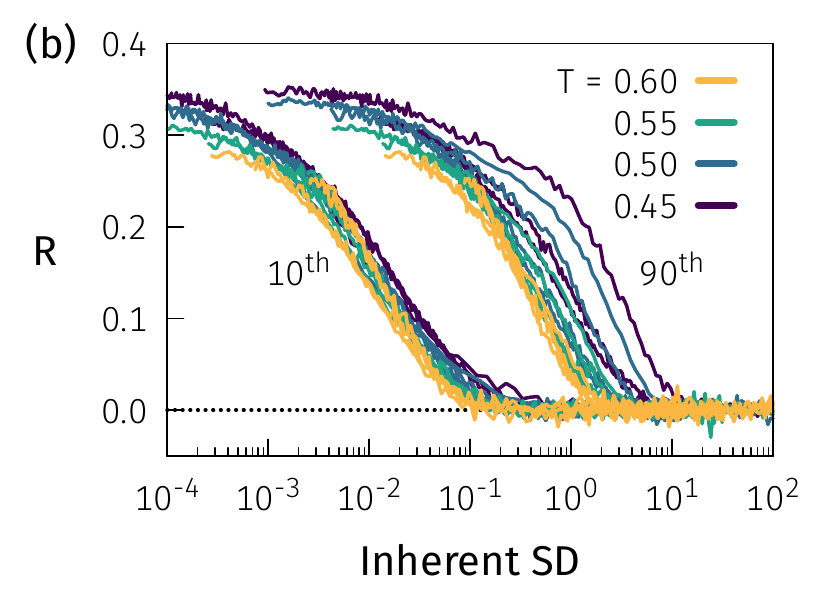}	
	\includegraphics[width=6.5cm]{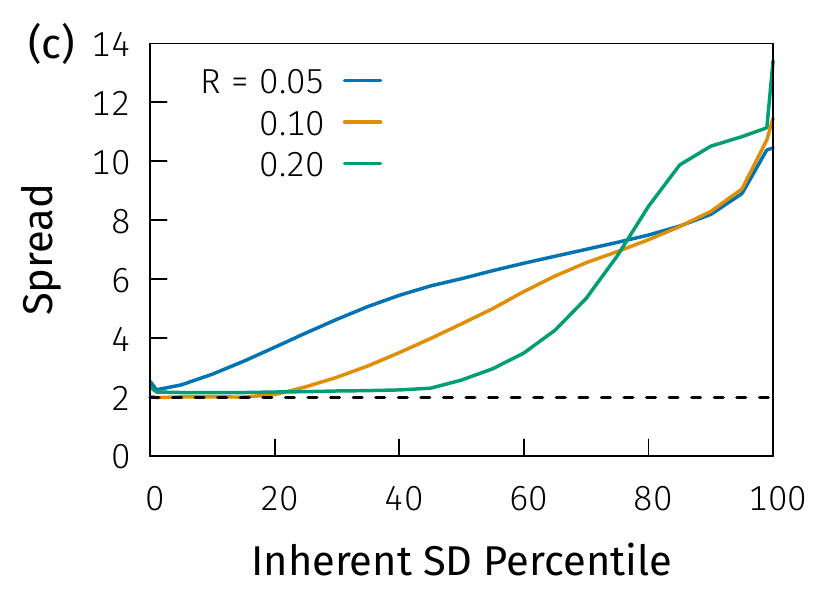}	
	\caption{\label{fig4} Slow-particle control of the dynamics. 
	(a) The slowest 25\% of particles as defined by the inherent displacement from an initial configuration on a time scale corresponding to the plateau region (equilibrium data at $T=0.50$). The cluster is seen to percolate the sample.
	(b) Normalized relaxation function for jumps to different temperatures (indicated by the color) as functions of the (inherent) slowest 10\textsuperscript{th} (left) and 90\textsuperscript{th} (right) percentile of particle displacement. The case of fewer slow particles collapses the curves better than when almost all particles are taken into account.
	(c) The spread of curves as in (b) as a function of the fraction of (inherent) slowest particles considered. The spread is measured at the three levels 0.05, 0.1, and 0.2. The spread obtained using the harmonic (inherent) MSD (\eq{harm}) at $R = 0.10$ is marked by the black dashed line.}
\end{figure}

As an illustration of the above, \fig{fig4}(a) shows a picture of the slowest 25\% of particles (determined by the inherent displacement over a time interval corresponding to the MSD plateau region). The slow particles percolate and maintain the overall structure to a large extent. It is only when the slow particles start to move and break up the percolating structure that significant relaxation takes place. This argument refers to equilibrium relaxation, but the same reasoning applies for aging following, e.g., a temperature jump. 

When a rigid structure percolates, one speaks of ``rigidity percolation'' as first discussed in connection with chalcogenide and other covalently bonded glasses \cite{phi85,bre86,tat90}. Inspired by this, the name ``dynamic rigidity percolation'' was introduced in 1989 to explain the frequency dependence of the sound velocity in a system of inverted micelles \cite{ye89}. This name signals what we propose control the physical aging. Note that that the percolating rigid structure changes with time as reflected by the fact that the collection of slow particles changes continuously. Enhanced packing or a low value of the ``softness'' \cite{sch17} leads to local structures that are more stable against rearranging than the surrounding, average liquid. In a flow situation, the rheological response is controlled by long-lived structures, i.e., the rearrangement and breaking of locally strong cages of nearest neighbors \cite{sti88,lau17} (see also Ref. \onlinecite{tra13} for thermodynamic consequences). 

How many slow particles should be included in the distance measure controlling physical aging? We considered a quarter in \fig{fig4}(a) because 25\% is a typical percolation threshold in three dimensions. \Fig{fig4}(b) investigates how well the normalized relaxation functions for different temperature jumps collapse as a function of the inherent MSD of the 10\% and 90\% slowest particles (the colors represent the annealing temperatures). We see that the 10\% fraction collapses the $R(t)$ data much better than the 90\% fraction. A systematic investigation is provided in \fig{fig4}(c), which reports the spread (width) of figures like those of (b) as a function of the fraction of slow particles. The spread depends on at which value the normalized relaxation functions are considered: blue corresponds to the spread at $R=0.05$, red to 0.1, and green to 0.2. In all cases, the fewer slow particles are considered, the smaller is the spread. This could be taken to suggest that the very slowest particle controls the overall aging. That would make little sense, however, and this is a wrong way of thinking about \fig{fig4}(c). It is more likely that the motion of the slowest particle is controlled by the overall structural relaxation than the other way around. In other words, once enough time has passed that there is little structural memory, even the slowest particles move.

The above does not answer the questions: How large a fraction of the slowest particles controls the aging? At what level of the normalized relaxation function $R(t)$ should one minimize the spread of curves like those in \fig{fig4}(b)? A pragmatic way of addressing this is to consider the (inherent) \textit{harmonic} MSD defined by averaging over all particles $i$ as follows \cite{ike13}

\be\label{harm}
\frac{1}{\langle\Delta\br^2\rangle_{\rm IHMSD}}
\,\equiv\,\left\langle\frac{1}{\Delta\br_i^2}\right\rangle_i\,.
\ee
This quantity is marked by the horizontal black dashed curve of \fig{fig4}(c), which is clearly a good choice for minimizing the spread and thus for a distance measure controlling the physical aging.

\begin{figure}[h]
	\includegraphics[width=7cm]{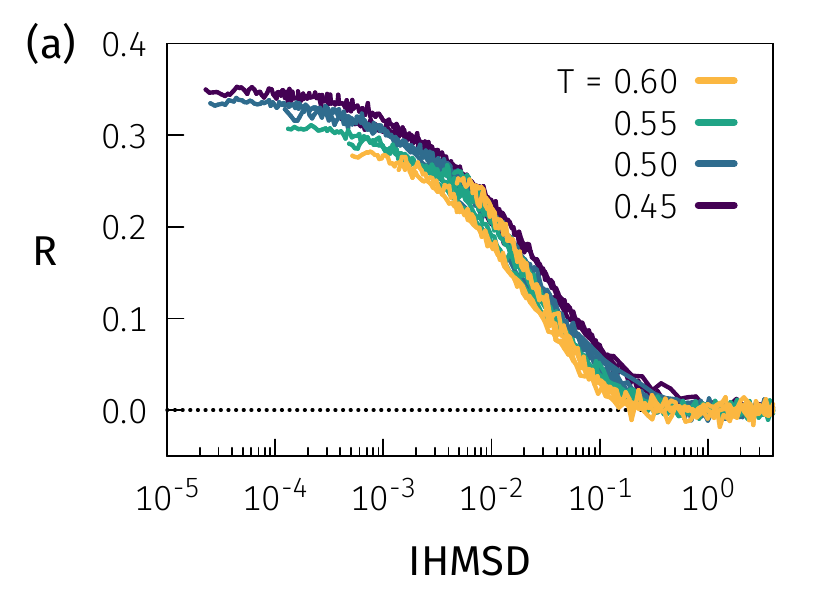}
	\includegraphics[width=7cm]{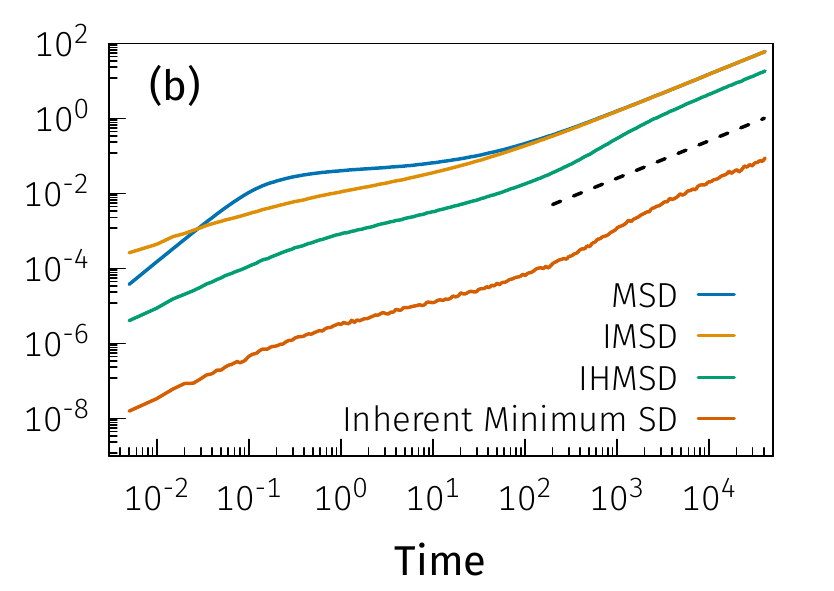}
	\caption{\label{fig5} Inherent harmonic MSD control of the material time.
	(a) The normalized relaxation function of the data of \fig{fig3}(a) plotted as a function of the inherent harmonic MSD. There is a good, but not perfect, overall collapse.
	(b)	Different distance measures in equilibrium at $T=0.50$ plotted as a function of time in a log-log plot. The blue curve is the standard MSD for which one observes a short-time ballistic regime of slope 2, a ``vibrational'' plateau at intermediate times, and a diffusive regime at large times, for which the MSD is proportional to time leading to a slope of unity as indicated by the black dashed line. The inherent MSD is the yellow curve that has no ballistic regime and follows the standard MSD at long times. The green curve is the inherent harmonic MSD (\eq{harm}) that uses the harmonic instead of the usual mean. Finally, the red curve is the inherent MSD of the single slowest particle, which has the same shape as the inherent harmonic MSD. At long times we find slope unity for all four curves, showing that distance measures based on any of them obey \eq{d12_limit}.}
\end{figure}

The conjecture that the inherent harmonic MSD controls aging is tested in \fig{fig5}(a), which plots the aging data as a function of $\langle\Delta\br^2\rangle_{\rm IHMSD}$. We see a nice, but not perfect collapse. The bifurcation of the responses at small displacements is greatly reduced for jumps to lower temperatures, becoming non-existent at $T = 0.50$. The spread of the normalized relaxation functions has been reduced from 3.5 decades (\fig{fig3}(b)) to 0.3 decades. There is, however, a systematic trend in the deviations from perfect collapse: Using a material time based on the inherent harmonic MSD, jumps to higher temperatures are consistently faster than those to lower temperatures. A possible cause of this could be that the system at higher temperatures can thermalize by moving a shorter distance than at lower temperatures, which would make sense since there are fewer states available at lower temperatures. A back-of-an-envelope calculation assuming a homogeneous distribution of states with density given by the inherent entropy, however, shows that this can only explain a spread of about 15\%, not the observed factor of two.

\Fig{fig5}(b) shows the different mean-square displacement measures at $T=0.50$: The standard MSD (blue), the inherent MSD (orange), the inherent harmonic MSD (green), and the inherent MSD of the single slowest particle. At long times these measures are all proportional to time, i.e., have the slope of unity required for a material-time to be defined via \eq{d12_limit}. Interestingly, the slowest particle MSD is very similar in shape to that of the inherent harmonic MSD. The same applies, e.g., for the slowest 10\% or 20\% (data not shown).

\section{Discussion}

\begin{figure}[h]
	\includegraphics[width=7cm]{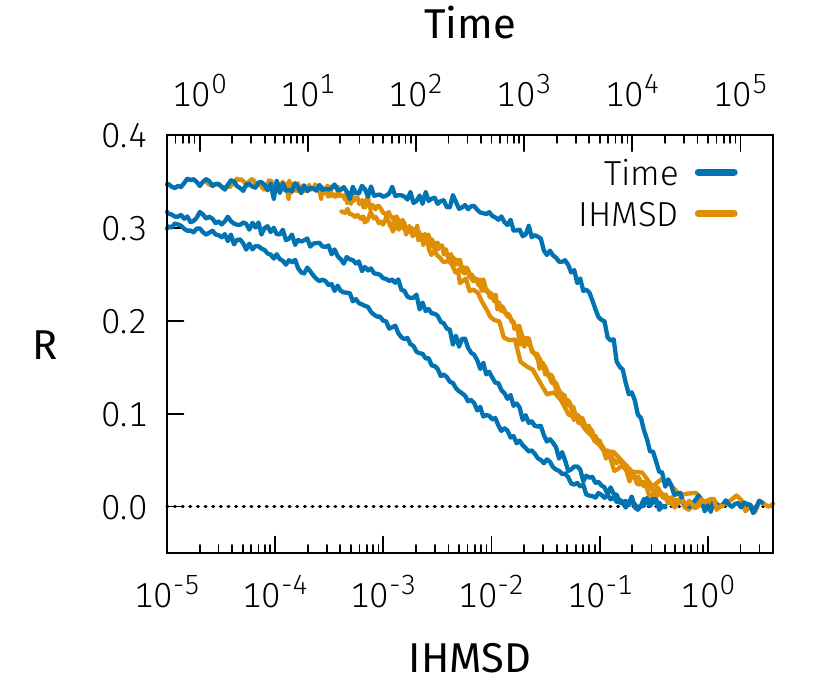}
	\caption{\label{fig6} The normalized relaxation function $R$ for all jumps to $T = 0.45$ as functions of time (blue) and of the proposed material-time candidate: the inherent harmonic MSD (orange). }
\end{figure}

We proceed to put the above findings into perspective by discussing their relation to related ideas; also, the TN equation is derived from minimal assumptions. First, however, we summarize this paper's main result in \fig{fig6} showing the normalized relaxation function limited to the jumps to the lowest annealing temperature studied ($T=0.45$). The relaxation functions are plotted as a function of time (blue) and of the inherent harmonic MSD (orange). Clearly, the latter quantity captures the essence of the physical aging.

\subsection{Relation to previous time-as-distance proposals}\label{disc}

As mentioned, it is not a new idea to quantify a time interval in terms of how far a system's particles have moved. This section discusses relevant previous works by first considering the linear-response case, followed by that of physical aging.

Linear-response theory is founded in the well-established fluctuation-dissipation (FD) theorem according to which the response is determined by a thermal-equilibrium time-correlation function \cite{reichl}. In the simplest case, that of a single energy bond  \cite{paynter,ost71,systemdyn,III}, the response is determined by an equilibrium time-autocorrelation function of some quantity $A$, denoted by $\langle A(t)A(t')\rangle$. For instance, if the externally controlled field is an electric field and the output is the current, $A$ is the current. Another familiar example is when the externally controlled field is the shear rate and the measured quantity is the shear stress, in which case $A$ is the shear stress. 

According to the FD theorem, the convolution kernel of the linear-response integral is the relevant equilibrium time-autocorrelation function. For a glass-forming liquid, the linear responses of different quantities slow down dramatically when the temperature is lowered, interestingly often in such a way that the ratio of their average relaxation times is independent of temperature \cite{jak12,roe21}. Another characteristic feature often observed is \textit{time-temperature superposition} (TTS). We proceed to argue that these characteristics find a natural explanation in the distance-as-time approach \cite{haa79,dyr97}.

The fact that any equilibrium time-autocorrelation function $\langle A(t_1)A(t_2)\rangle\,\,\,(t_1<t_2)$ goes to zero as $t_2-t_1\to\infty$ has the simple ``geometric'' interpretation that the memory of a system's properties at time $t_1$ fades as the particles move away from their positions at $t_1$. Because of this, an obvious idea is to regard the convolution kernel of linear-response theory as a function of the distance traveled in configuration space \cite{dyr97}. This line of thinking arose in a paper by Haan from 1979 \cite{haa79}, who proposed to describe particle motion in a liquid by means of a Smoluchowski equation in which the role of time is played by the MSD over the time interval in question. 

In thermal equilibrium the distance traveled is in a one-to-one correspondence with the time that has passed (\sect{eq}), implying that the geometric viewpoint must be correct in the sense that any time-autocorrelation function  $\langle A(0)A(t)\rangle$ is a unique function of the distance traveled in time $t$. If the corresponding ``geometric'' equilibrium autocorrelation function is denoted by $\phi_A(d)$, this means that $\langle A(0)A(t)\rangle=\phi_A(d(t))$ in which $d(t)$ is the distance between two points along the equilibrium trajectory the time $t$ apart \cite{dyr97}. As formulated here, this is a tautology. Nevertheless, it provides a simple rationalization of TTS because TTS will apply for any quantity $A$ if 1) the MSD obeys TTS, and 2) $\phi_A(d)$ is temperature independent. The fact that different linear-response quantities slow down in a concerted fashion \cite{jak12,roe21} follows if the relevant geometric autocorrelation functions are temperature independent, in which case the slowing down is controlled entirely by the slowing down of the particle motion. Two interesting special cases are the following \cite{dyr97}. 1) If there is a Gaussian long-distance decay of the ``geometric'' equilibrium autocorrelation function, i.e., if $\ln\left(\phi_A(d)\right)\propto -R^2$ for $R\to\infty$ where $R$ is the geometric distance, the time-autocorrelation function has a long-time simple exponential decay because $R^2(t)\propto t$ as $t\to\infty$. Such a decay corresponds to a Debye frequency-dependence of the linear response, which in the frequency domain is often observed below the relevant loss peak \cite{hec18,roe21}. 2) If there is an exponential long-distance decay of the geometric equilibrium autocorrelation function, i.e., $\ln\left(\phi_A(d)\right)\propto -R$ for $R\to\infty$ \cite{wee07,bai09}, the long-time behavior corresponds to a stretched exponential with exponent $1/2$, which fits many experiments well \cite{dyr06a,dyr07,nie09}. 

Turning now to aging, as previously mentioned the main ingredients of \sect{pa} are present in pioneering papers by Cugliandolo, Kurchan (CK), and coworkers from the 1990s \cite{cug94,cha02,cug97}. These authors developed a theory of physical aging of the infinite range Sherrington-Kirkpatrick spin glass based on Schwinger–Dyson equations for both correlations and susceptibility, utilizing the fact that mean-field theory is exact for this model \cite{cha07,cug11}. The results obtained, which generalize to finite-range spin glasses \cite{cha07}, are based on a demonstration of time-scale separation and time-reparametrization invariance of the effective dynamical action describing the slow degrees of freedom; we note that the latter symmetry was recently applied to models of black holes and ``strange'' quantum liquids \cite{fac19}. The physical idea is that the proper measure of the time interval between $t_1$ and $t_2$ is not the ``wall clock in the laboratory'', $t_2-t_1$, but the value of the spin autocorrelation function $\langle S(t_1)S(t_2)\rangle$ \cite{cha07}. For Ising spins with values $\pm 1$, the spin autocorrelation function of two configurations is in a one-to-one relation to the Euclidean distance squared between the two configurations. This means that, just as above, time intervals are quantified in terms of the distance in configuration space. Moreover, time-reparametrization invariance implies a ``triangular relation'' according to which for times $t_1<t_2<t_3$ the two spin autocorrelation functions $\langle S(t_1)S(t_2)\rangle$ and $\langle S(t_2)S(t_3)\rangle$ determine $\langle S(t_1)S(t_3)\rangle$. Because the spin autocorrelation function determines the distance, this is equivalent to the unique-triangle property \eq{utp}. The analog of the above discussed geometric reversibility of aging (the second equality sign of \eq{utp}) was also derived by CK in the form of commutativity of an algebraic relation defined from the spin autocorrelation functions \cite{cug94}. 

Despite these similarities between the ideas of the present paper and those of CK, our emphasis is different. CK introduced the concept of a waiting time $t_w$ and made the important discovery that the spin autocorrelation function factorizes as follows $C(t_w,t_w+\tau)=F[h(t_w)/h(t_w+\tau)]$ for some functions $F$ and $h(t)$; this is equivalent to \eq{xid12} if the material time is controlled by a distance and it is assumed that the autocorrelation function is determined by the difference in material time. CK focused on quenching a system from thermal equilibrium to a low temperature, not on how the system approaches equilibrium. In contrast, that is precisely the focus of the material-time description, which from the outset was devised for relatively small temperature perturbations. In fact, the TN description often breaks down for larger perturbations, both in experiment \cite{scherer,mck17,rie22} and in simulations \cite{kob00}, leading to an interesting question for future work: Why does a material-time description work well both for modest perturbations of equilibrium (TN) and for extreme perturbations (CK) -- but often not in the intermediate regime? Related to this question is an additional important difference between the CK and present approaches: References \onlinecite{cug94,cug97,cas07,cha07,cug11} emphasize that spatial heterogeneities lead to local clocks ticking with different rates: ``a region looks older than another one when observed on a given time window'' \cite{cha07}. In contrast, we define a single global material time and take dynamic heterogeneities into account by assuming that the material time is controlled by the slowest moving particles.

Schober studied the aging of pressure and potential energy of the Kob-Andersen (KA) binary LJ system and found that these quantities age ``in parallel'', i.e., have the same normalized relaxation functions \cite{sch12} (this result, incidentally, follows from the fact that the KA system has strong virial potential-energy correlations \cite{ped08,I,IV,ped18}). He showed that both quantities age following an exponential function of the single-particle MSD minus the vibrational MSD, which is close to the above discussed inherent MSD. He also found that pressure and energy follow the drop in diffusivity. A few years later, in a study of the breakdown of the Stokes-Einstein relation between viscosity and diffusion coefficient, Schober and Peng \cite{sch16} proposed to use the van Hove self-correlation function to distinguish between slow and fast particles because the latter are seen mainly in the non-Gaussian tail of the displacement distribution. The contribution of slow particles to the viscosity increases upon cooling, a result that demonstrates their importance for the highly viscous liquid phase and which is consistent with the above discussed dynamic-rigidity-percolation picture. Recently, Schober studied numerically the physical aging of liquid and amorphous selenium \cite{sch21} and found that also for this system, the pressure and potential energy age with a relaxation function that at long times is described by an exponential function of the MSD minus the vibrational MSD.

\subsection{The Tool-Narayanaswamy description}\label{tn}

This section proposes an Occam's-razor type justification of the TN equation \eq{TN_resp}. In the process we relate to TTS and to the experimental fact that different linear-response functions often have the same temperature dependence. Note that TTS follows rigorously from the TN formalism, i.e., that TTS is a necessary condition for TN to apply.

We start by expressing linear-response theory in terms of a yet unspecified distance measure $d$ obeying \eq{d12_limit}. If the externally controlled input is $e_b(t)$ and the output is $q_a(t)$, standard linear-response theory is expressed by the Stieltjes integral (assuming $\langle q_a\rangle=0$)

\be\label{lin_resp}
q_a(t)
\,=\,\int_{-\infty}^t\phi_{ab}(t-t')\,\delta e_b(t')\,.
\ee
Here, as is well known, \textit{linearity} is reflected by the fact that only the first power of the field appears, \textit{causality} by the fact that $q_a(t)$ only depends on $\delta e_b(t')$  for $t'<t$, and \textit{time-translational invariance} by the fact that the convolution kernel $\phi_{ab}$ only depends on the time difference $t-t'$. 

Whenever \eq{d12_limit} applies, $t-t'$ in \eq{lin_resp} may be replaced by a distance measure $d$ by proceeding as follows. If one picks a fixed time in the far distant past, $t_0$, \eq{d12_limit} implies that two constants $\alpha$ and $\beta$ exist such that 

\be\label{dabeq}
d(t_0,t)
\,=\,\alpha t + \beta\,.
\ee 
This means that $t-t' = \left[d(t_0,t)-d(t_0,t')\right]/\alpha$. When substituted into \eq{lin_resp} this leads to (with $\psi_{ab}(x) \equiv \phi_{ab}(x/\alpha)$) 

\be\label{eq3a}
q_a(t)
\,=\,\int_{-\infty}^t\,\psi_{ab}(d(t_0,t)-d(t_0,t'))\,\delta e_b(t')\,.
\ee
\Eq{eq3a} is merely a reformulation of linear-response theory and, as such, rigorously obeyed for any sufficiently small perturbation applied to a state of thermal equilibrium. Note that any distance measure $d$ obeying \eq{d12_limit} may be used in \eq{eq3a}. 

Next we \textit{assume} that \eq{eq3a} not just applies in equilibrium at the single temperature $T$, but also for aging systems subject to temperatures varying around $T$. Whether or not this is a realistic assumption depends, of course, on the choice of the distance measure. This assumption has three consequences:

\begin{enumerate}

\item \textit{TTS for the equilibrium linear response}\\
In equilibrium at temperature $T$, \eq{dabeq} and \eq{eq3a} lead to

\be\label{eq4}
q_a(t)
\,=\,\int_{-\infty}^t \psi_{ab}(\alpha(T)(t-t'))\,\frac{d e_b(t')}{dt'}\,dt'\,.
\ee
Writing $q_a(t)=q_a(\omega)\exp(i\omega t)$ and $e_b(t)=e_b(\omega)\exp(i\omega t)$, \eq{eq4} implies $q_a(\omega)=R_{ab}(\omega)e_b(\omega)$ in which

\be\label{R_eq}
R_{ab}(\omega)
\,=\,i\omega\int_0^\infty \psi_{ab}(\alpha(T)t'')\,e^{-i\omega t''}\,dt''\,.
\ee
This implies that $R_{ab}$ is a function of $i\omega/\alpha(T)$, i.e., TTS. 

\item \textit{Proportional time scales for different linear-response functions in thermal equilibrium}\\
It follows from the above that the temperature dependence of the response function $R_{ab}(\omega)$ is determined by $\alpha(T)$. This number depends only on how fast the particles move in equilibrium at temperature $T$, not on the generalized charge $q_a$ or the external field $e_b$. Consequently, the time scales of  different linear-response functions are the same. This means that all linear-response functions controlled by geometry in the above sense must have the same temperature dependence of their characteristic relaxation times, i.e., a temperature-independent ratio of their average relaxation times \cite{jak12,roe21}.

\item \textit{The TN formalism}\\
Defining the material time by

\be\label{xi}
\xi(t)
\,\equiv\, d(t_0,t)\,,
\ee
and substituting into \eq{eq3a}, one arrives at the TN equation \eq{TN_resp}.
\end{enumerate}

To summarize, if the material time via \eq{xi} is defined in terms of a distance measure obeying \eq{d12_limit} in which $t_0$ is a time in the far distant past, the TN formalism follows if the equation describing the equilibrium linear response also applies for out-of-equilibrium situations. This ``derivation'' may explain why the TN formalism usually works best for relatively small temperature variations \cite{scherer,mck17,rie22}. Thus our suggestion for why the TN formalism works well in the latter situation is that this regime is \textit{pseudolinear} in the sense that the geometry of particle motion is the same as that of thermal equilibrium, which is indeed the finding of \fig{fig2}(d). From this point of view, it is only when the ``wrong'' time variable is used -- the laboratory time -- that physical aging is strongly nonlinear and violates time-translational invariance.

\section{Summary}

Although the TN formalism for the description of physical aging has been around for half a century and is used routinely in industry, there have been few attempts to justify it theoretically. We propose that the inherent harmonic mean-square displacement, which emphasizes the role of the slowest particles, is the quantity that controls the material time according to a dynamic-rigidity-percolation picture. The idea is that little overall relaxation takes place as long as a percolating structure of particles, which have barely moved, is maintained. None of the ingredients of our approach are new, compare \sect{disc}, but they have here been combined with a focus on defining the material time and explaining the origin of the TN linear convolution integral \eq{TNeq}. We hope that the considerations presented in this paper may inspire to works aimed at further elucidating the physical origin of the material time and why TN works so well.

\acknowledgments{We would like to thank Camille Scalliet for suggesting the harmonic inherent mean-square displacement as a useful means to emphasize the slow-particle displacements. This work was supported by the VILLUM Foundation's \textit{Matter} grant (16515).}

\end{document}